\documentclass[lettersize,journal]{IEEEtran}

\usepackage{amsmath,amsfonts}
\allowdisplaybreaks
\usepackage{amssymb}
\usepackage{amsthm}
\usepackage{upgreek}
\usepackage{algorithmic}
\usepackage{algorithm}
\usepackage{array}
\usepackage{subcaption}
\usepackage{textcomp}
\usepackage{stfloats}
\usepackage{url}
\usepackage[table]{xcolor}
\usepackage{verbatim}
\usepackage{graphicx}
\usepackage{cite}
\usepackage{booktabs}
\usepackage{hyperref}
            
\usepackage{tikz}
\usetikzlibrary{positioning, arrows.meta, calc}

\usepackage{multirow}
\usepackage{siunitx}

\usepackage{acronym}
\newacro{ACOPF}{Alternating Current Optimal Power Flow}
\newacro{CCT}{Critical Clearing Time}
\newacro{COI}{Center of Inertia}
\newacro{DAE}{Differential-Algebraic Equation}
\newacro{IBG}{Inverter-Based Generator}
\newacro{KKT}{Karush-Kuhn-Tucker}
\newacro{LMP}{Locational Marginal Price}
\newacro{OPF}{Optimal Power Flow}
\newacro{SG}{Synchronous Generator}
\newacro{TS}{Transient Stability}
\newacro{TSCOPF}{Transient Stability Constrained Optimal Power Flow}

\begin{document}

\title{Shadow Pricing in Transient Stability-Constrained Optimal Power Flow}

\author{Alex Coelho, \IEEEmembership{Student Member,~IEEE}, Araceli Hernandez, \IEEEmembership{Senior Member,~IEEE}, and Luis Badesa
}



\maketitle

\begin{abstract}
Transient stability constraints alter the economic dispatch of power systems, while the resulting price signals and their implications for market settlement remain insufficiently characterized. Existing approaches identify stability-related components in locational marginal prices, without deriving shadow prices for individual assets, such as generators and loads, or analyzing the resulting settlement. To address this gap, this paper derives shadow prices from a transient stability-constrained optimal power flow in which the differential-algebraic equations governing the system dynamics are embedded through direct discretization. Each shadow price decomposes into a nodal energy component and a stability component, the latter formed by the dual variables of the dynamic equations in which the power of the asset appears. Shadow prices are therefore asset-specific: co-located generators and loads are assigned the same nodal energy component, but can be subject to different stability components when their transient behaviors differ. Case studies on the WECC 9-bus system and a modified IEEE 39-bus system show that settlement at the proposed shadow prices allocates the cost of the stability constraints among the assets without any side payment. An asset that reduces the stability margin is charged its stability component, whereas an asset that improves it is paid, and no generator is dispatched at a loss.
\end{abstract}

\begin{IEEEkeywords}
Transient stability, optimal power flow, shadow prices, market settlement
\end{IEEEkeywords}

\section{Introduction}
\label{sec:intro}

\IEEEPARstart{P}{ower} systems are increasingly operated close to their stability limits due to the growth of renewable generation, the retirement of conventional plants, and the pressure of competitive electricity markets~\cite{benidris_2017_intro}. Under these conditions, \ac{TS} may be compromised, i.e., \acp{SG} and \acp{IBG} may fail to remain in synchronism following a large disturbance, potentially leading to blackouts and significant economic and societal impacts~\cite{IEEE_TF_2021}. A least-cost dispatch that does not account for \ac{TS} may therefore have to be corrected through generation redispatch to preserve synchronism after a contingency, potentially resulting in significant operating costs~\cite{kheradmandi2011rescheduling}. \ac{TSCOPF} addresses this issue by incorporating \ac{TS} constraints into the dispatch problem, combining the economic objective and steady-state network equations of a conventional \ac{OPF} with constraints that ensure stable post-fault behavior, i.e., generators remaining in synchronism, for a set of credible contingencies~\cite{zhang_2025}. However, when a stability constraint affects the optimal dispatch, it also affects the shadow prices used to settle the market. This raises a fundamental question that has received comparatively little attention: \emph{what is the economic value of \ac{TS} services, and how do the resulting shadow prices affect the market?}

\IEEEpubidadjcol

A recent survey~\cite{zhang_2025} traces \ac{TSCOPF} research back to the 1980s and identifies four main solution approaches. Discretization methods embed the \acp{DAE} governing the transient trajectory directly into the optimization problem~\cite{aghahassani2024}, whereas sequential approaches alternate between time-domain simulation and optimization~\cite{pizanomartinez_2014}. Heuristic methods seek feasible solutions through predefined search rules~\cite{wang_2020_heuristic}, while machine-learning approaches use data-driven surrogates to replace dynamic simulations or approximate stability constraints~\cite{wu_2024_nn_uctscopf, garcia2025}. Across these approaches, recent research has primarily focused on improving the fidelity, scalability, and computational performance of TSCOPF formulations. For example, higher-order generator and converter models have been incorporated to represent a broader range of dynamic behavior~\cite{liederer2022,aghahassani2024}, while studies on numerical integration, load modeling, and decomposition methods have investigated their effects on convergence, computational burden, and operating cost~\cite{aghahassani_2022_nm,arredondo2023,wu_2024_nn_uctscopf}. Despite these developments, the economic analysis of the resulting \ac{TS} constraints has generally remained limited to comparing the total operating cost of \ac{TSCOPF} relative to a conventional \ac{OPF}.

This predominant focus on the primal problem contrasts with the more mature literature on the economic valuation of frequency stability services~\cite{badesa_2023_assigning, matamala_2024_cost_allocation}. In the \ac{TS} domain, few studies have examined how stability constraints affect \acp{LMP} or explored how these constraints can be assigned an explicit monetary value. Yuan \emph{et al.}~\cite{yuan2004} proposed one of the earliest formulations by introducing a system-wide stability charge based on the difference between the maximum rotor-angle excursion and an operator-defined threshold. Although conceptually establishing \ac{TS} as a priced security attribute, the proposed charge was not numerically evaluated. Subsequent studies moved toward locational formulations. Amarasinghe \emph{et al.}~\cite{amarasinghe_2005} and Annakkage \emph{et al.}~\cite{annakkage2006} used kernel ridge regression to approximate the \ac{CCT} based on steady-state voltage phasors, allowing a differentiable stability constraint to be incorporated into an optimization-based pricing framework. They showed that the Lagrange multiplier of the stability constraint contributes to the \acp{LMP} and reported a corresponding increase in network rental. However, the priced quantity remains the multiplier of a learned approximation of the \ac{CCT}, and the reported increase in network rental is not traced to the individual generators and loads.

Nguyen \emph{et al.}~\cite{nguyen2011} used the active power of each generator and some control parameters to approximate linear functions of maximum rotor-angle deviation. They then incorporated these approximated functions as \ac{TS} constraints into the \ac{OPF}. This approach enabled them to identify a \ac{TS} component within the generator-bus \acp{LMP}. However, the linearized constraints could represent only the operating point at which the sensitivities were obtained. This demanded a new iterative procedure whenever a new operating point was to be analyzed. More recently, machine-learning approaches have offered alternative representations of the stability constraint. Garcia \emph{et al.}~\cite{garcia2025} embedded a neural-network stability classifier directly into the \ac{OPF} and derived shadow prices associated with generators from the corresponding Lagrange multipliers and classifier gradients. Depending on the features used by the classifier, the resulting prices can be either purely locational or can differ between generators connected to the same bus. However, their work focuses on the accuracy of the neural-network representation of the stability region. The generator-specific prices are derived, characterized as discriminatory, and left out of the numerical results. In addition, the shadow prices associated with loads are not formulated, the effect of the load behavior on the stability constraints is not considered, and the resulting settlement is not analyzed.

Taken together, these studies demonstrate that stability constraints can influence \acp{LMP}, but they differ substantially in the \emph{stability metric being priced, the mathematical representation of the stability constraint, and the economic interpretation of its Lagrange multiplier}. More importantly, the existing literature does not provide a formulation of the stability component of the shadow price that follows from \ac{TSCOPF} with directly discretized \acp{DAE}, in which the transient dynamics are embedded in the optimization problem without a surrogate. Nor does it establish how such a component should be interpreted and settled for individual generators and loads. To address these gaps, this paper formulates \ac{TSCOPF} based on the discretization of the \acp{DAE}, which accommodates multiple dynamic model components, and derives the shadow prices associated with loads and generators from its dual variables.

Considering the identified gaps in the literature, the main contributions of this work are the following:

\begin{enumerate}
    \item To derive shadow prices for generators and loads in \ac{TSCOPF} by directly discretizing the \acp{DAE}, which makes the stability component of each shadow price follow from the time-domain equations enforced in the dispatch, regardless of the dynamic model adopted for each asset.
    \item To show that the stability component is specific to each asset, making co-located generators and loads subject to different shadow prices. Under a single \ac{LMP}, assets connected at the same bus would not receive the appropriate compensation.
    \item To demonstrate that settlement at the proposed shadow prices allocates the cost of securing stability among the assets without a separate allocation rule, charging those that reduce the stability margin and remunerating those that improve it.
\end{enumerate}

The remainder of this paper is organized as follows. Section~\ref{sec:tscopf} presents \ac{TSCOPF} in discretized form. Section~\ref{sec:economics_tscopf} derives the shadow prices associated with loads and generators from the dual variables of that formulation, defines their stability components, and presents the resulting settlement. Section~\ref{sec:case_studies} applies the formulation to the WECC 9-bus system and a modified IEEE 39-bus system. Finally, Section~\ref{sec:conclusions} concludes the paper and outlines future research.

\section{Transient Stability Constrained \ac{OPF}}
\label{sec:tscopf}

\ac{TSCOPF} extends \ac{ACOPF} with the transient dynamics of the power system following a contingency. The steady-state operating point given by the dispatch sets the initial conditions of these dynamics, and the angles of the generators during the transient are required to satisfy a prescribed stability criterion. The dispatch is therefore optimized jointly with the transient response it produces. In this paper, the dynamic equations are embedded in the optimization problem through time discretization.

\subsection{Steady-State \ac{ACOPF}}
\label{subsec:acopf}

Consider a power system with sets of buses $\mathcal{B}$, branches $\mathcal{L}$, generators $\mathcal{G}$ and loads $\mathcal{D}$. The subsets $\mathcal{G}_k\subseteq\mathcal{G}$ and $\mathcal{D}_k\subseteq\mathcal{D}$ collect the generators and loads connected at bus $k$, and the subscripts $g(k)$ and $d(k)$ denote generator $g\in\mathcal{G}_k$ and load $d\in\mathcal{D}_k$. Bus voltages are expressed in polar coordinates through the magnitude $V_k$ and the angle $\theta_k$, and the network through the bus admittance matrix $\boldsymbol{\mathrm{Y}}=\boldsymbol{\mathrm{G}}+\jmath\boldsymbol{\mathrm{B}}$. The net power injections at bus $k$ are
\begin{align}
P_k &= V_k\sum_{m\in\mathcal{B}} V_m
\left(\mathrm{G}_{km}\cos\theta_{km}
+\mathrm{B}_{km}\sin\theta_{km}\right)
\label{eq:p_injection}\\
Q_k &= V_k\sum_{m\in\mathcal{B}} V_m
\left(\mathrm{G}_{km}\sin\theta_{km}
-\mathrm{B}_{km}\cos\theta_{km}\right)
\label{eq:q_injection}
\end{align}
where $\theta_{km}=\theta_k-\theta_m$. The steady-state operating point is obtained from the conventional \ac{ACOPF},
\begin{subequations}
\label{eq:acopf}
\begin{alignat}{2}
& \hspace*{-12mm} \min\;\;
  \sum_{g\in\mathcal{G}} \mathrm{c}_g P_{g}
\label{eq:obj}\\[-1pt]
& \hspace*{-11.5mm} \text{subject to:} \notag\\[-2pt]
(\lambda_{k})\quad
& P_{k}
  + \sum_{d\in\mathcal{D}_k}\mathrm{P}_{d(k)}
  - \sum_{g\in\mathcal{G}_k}P_{g(k)} = 0
  &&\ \forall\,k\in\mathcal{B}
\label{eq:pbalance}\\
(\beta_{k})\quad
& Q_{k}
  + \sum_{d\in\mathcal{D}_k}\mathrm{Q}_{d(k)}
  - \sum_{g\in\mathcal{G}_k}Q_{g(k)} = 0
  &&\ \forall\,k\in\mathcal{B}
\label{eq:qbalance}\\
(\zeta_{\mathrm{ref}})\quad
& \theta_{k} = 0
  &&\ k = \text{ref}
\label{eq:ref} \\
(\overline{\mu}_{g})\quad
& P_{g}
  - \mathrm{P}^{\max}_{g} \leq 0
  &&\ \forall\,g\in\mathcal{G}
\label{eq:pgmax}\\
(\underline{\mu}_{g})\quad
& \mathrm{P}^{\min}_{g}
  - P_{g} \leq 0
  &&\ \forall\,g\in\mathcal{G}
\label{eq:pgmin}\\
(\overline{\eta}_{g})\quad
& Q_{g}
  - \mathrm{Q}^{\max}_{g} \leq 0
  &&\ \forall\,g\in\mathcal{G}
\label{eq:qgmax}\\
(\underline{\eta}_{g})\quad
& \mathrm{Q}^{\min}_{g}
  - Q_{g} \leq 0
  &&\ \forall\,g\in\mathcal{G}
\label{eq:qgmin}\\
(\overline{\alpha}_{k})\quad
& V_{k}
  - \mathrm{V}^{\max}_{k} \leq 0
  &&\ \forall\,k\in\mathcal{B}
\label{eq:vmax}\\
(\underline{\alpha}_{k})\quad
& \mathrm{V}^{\min}_{k}
  - V_{k} \leq 0
  &&\ \forall\,k\in\mathcal{B}
\label{eq:vmin}\\
(\gamma_{km})\quad
& P_{km}^{\,2} + Q_{km}^{\,2}
  - \bigl(\mathrm{S}^{\max}_{km}\bigr)^{2} \leq 0\rlap{$\quad\forall\,(km)\in\mathcal{L}$}
  &&
\label{eq:sflim}\\
(\gamma_{mk})\quad
& P_{mk}^{\,2} + Q_{mk}^{\,2}
  - \bigl(\mathrm{S}^{\max}_{km}\bigr)^{2} \leq 0\rlap{$\quad\forall\,(mk)\in\mathcal{L}$}
  &&
\label{eq:stlim}
\end{alignat}
\end{subequations}
where \eqref{eq:obj} minimizes the total generation cost, \eqref{eq:pbalance} and \eqref{eq:qbalance} enforce the nodal active and reactive power balance, \eqref{eq:ref} fixes the angle at the reference bus, and \eqref{eq:pgmax}--\eqref{eq:stlim} impose the generation, voltage and branch-flow limits. Lagrange multipliers are given in parentheses to the left of each constraint.

\subsection{Discretized \acp{DAE}}
\label{subsec:dynamic_model}

The transient dynamics enter the optimization problem by discretizing the \acp{DAE} over the set of uniformly spaced time instants $\mathcal{T}=\{t_1, t_2, \ldots,t_n\}$ with step $\Delta t$. Each dynamic component, e.g., a generator and its controls, contributes its initialization and its discretized \acp{DAE} as equality constraints for the \ac{TSCOPF}. Let $\boldsymbol{\mathrm{x}} = \left[\boldsymbol{V}_{\mathcal{B}}^\top, \boldsymbol{\theta}_{\mathcal{B}}^\top, \boldsymbol{P}_{\mathcal{G}}^\top, \boldsymbol{Q}_{\mathcal{G}}^\top\right]^\top$ be the vector of steady-state variables and $\boldsymbol{\mathrm{u}}$ the model input parameters. Vector $\boldsymbol{\mathrm{y}}^0$ holds the initial conditions and $\boldsymbol{\mathrm{y}}^t$ contains the relevant dynamic and algebraic variables at each time step, including generator voltage, angle, frequency and output power. This gives
\begin{subequations}
\label{eq:dynamic_constraints}
\begin{align}
    (\boldsymbol{\tau}^0)\quad&
    \boldsymbol{\mathrm{y}}^0 - \boldsymbol{f}^0(\boldsymbol{\mathrm{x}},\boldsymbol{\mathrm{u}})=0
    \label{eq:init_link}\\
    (\boldsymbol{\tau}^t)\quad&
    \boldsymbol{\mathrm{y}}^t - \boldsymbol{f}^t(\boldsymbol{\mathrm{y}}^{t}, \boldsymbol{\mathrm{y}}^{t-1},\boldsymbol{\mathrm{u}})=0
    \qquad \forall \; t \in \mathcal{T}
\label{eq:dae_step}
\end{align}
\end{subequations}
where $\boldsymbol{\tau}^0$ and $\boldsymbol{\tau}^t$ are the associated Lagrange multipliers. Map $\boldsymbol{f}^0$ initializes each dynamic component at the steady-state operating point, so \eqref{eq:init_link} ties the dispatch $\boldsymbol{\mathrm{x}}$ to the transient dynamics. In \eqref{eq:dae_step}, $\boldsymbol{f}^t$ is the implicit discretization of the \acp{DAE} and couples $\boldsymbol{\mathrm{y}}^t$ to the previous step $\boldsymbol{\mathrm{y}}^{t-1}$. Enforced at every $t\in\mathcal{T}$, this coupling propagates the initial conditions set by $\boldsymbol{\mathrm{x}}$ along the whole window.

The formulation accommodates \acp{SG} with their control systems, such as exciters, turbine-governors, and power system stabilizers, as well as other dynamic components represented by \ac{DAE} models. The proposed formulation and shadow-price derivation are independent of the particular dynamic models adopted, provided that their dynamics can be represented by discretized equations. For the case studies in Section \ref{sec:case_studies}, the transient dynamics are discretized using the trapezoidal integration method, with a fourth-order \ac{SG} model, an AVR of type SEXS, and a turbine-governor of type TGOV1. The corresponding initialization and dynamic equations follow \cite{liederer2022}.

The derivation of the shadow price associated with generators in Section~\ref{subsec:price_gen} uses only the subset of \eqref{eq:dynamic_constraints} in which the steady-state active power generated ($P_g$) appears explicitly. This subset is relevant because marginal changes in $P_g$ directly modify the initial active-power balance of the generators and, consequently, the accelerating power that drives the angle dynamics. These constraints couple $P_g$ to the initial electrical and mechanical powers of the dynamic model, linking the steady-state dispatch to the initialization of the generator dynamics
\begin{subequations}
\label{eq:initial_coupling}
\begin{align}
(\sigma_{\mathrm{P},g}^0)\quad&
P_g-P_{\mathrm{e},g}^0=0
\label{eq:init_Pe}\\
(\chi_{\mathrm{m},g}^0)\quad&
P_g-P_{\mathrm{m},g}^0=0
\label{eq:init_Pm}
\end{align}
\end{subequations}

Other generation technologies follow the same structure, including \acp{IBG} \cite{aghahassani2024}, whose dynamic model supplies its own initialization and discretized \acp{DAE}.

\subsection{Transient Network Equations}
\label{subsec:dynamic_network}

The network is kept in full nodal form at every step of the fault-on and post-fault windows. Demand follows a ZIP model referred to the pre-fault voltage $V_{k}$, with coefficients $(\mathrm{a_{Z}},\mathrm{a_{I}},\mathrm{a_{P}})$ for active power and $(\mathrm{b_{Z}},\mathrm{b_{I}},\mathrm{b_{P}})$ for reactive power. The power balances $\forall \; k \in \mathcal{B}$ and $\forall \; t \in \mathcal{T}$ are
\begin{subequations}
\label{eq:net_dyn}
\begin{align}
(\hat{\lambda}^{t}_{k}) \;
& V^{t}_{k}\sum_{m\in\mathcal{B}}V^{t}_{m}
  \left(\mathrm{G}^{\mathrm{f/pf}}_{km}\cos\theta^{t}_{km}
       +\mathrm{B}^{\mathrm{f/pf}}_{km}\sin\theta^{t}_{km}\right)
  \nonumber\\
& + \sum_{d\in\mathcal{D}_k}\mathrm{a}^{t}_{d(k)}\,\mathrm{P}_{d(k)}
  - \sum_{g\in\mathcal{G}_k}P^{t}_{\mathrm{e},g(k)} = 0
\label{eq:net_p_ts}\\
(\hat{\beta}^{t}_{k}) \;
& V^{t}_{k}\sum_{m\in\mathcal{B}}V^{t}_{m}
  \left(\mathrm{G}^{\mathrm{f/pf}}_{km}\sin\theta^{t}_{km}
       -\mathrm{B}^{\mathrm{f/pf}}_{km}\cos\theta^{t}_{km}\right)
  \nonumber\\
& + \sum_{d\in\mathcal{D}_k}\mathrm{b}^{t}_{d(k)}\,\mathrm{Q}_{d(k)}
  - \sum_{g\in\mathcal{G}_k}Q^{t}_{\mathrm{e},g(k)} = 0
\label{eq:net_q_ts}
\end{align}
\end{subequations}
where $\mathrm{G}_{km}^{\mathrm{f/pf}}$ and $\mathrm{B}_{km}^{\mathrm{f/pf}}$ correspond to the fault-on or post-fault topology, $\theta_{km}^{t}=\theta_k^t-\theta_m^t$, and the voltage-dependent ZIP factor of load $d$ at bus $k$ is
\begin{equation}
\mathrm{a}^{t}_{d(k)}=\mathrm{a}_{\mathrm{Z},d}
 \left(\frac{V^{t}_{k}}{V_{k}}\right)^{2}
 +\mathrm{a}_{\mathrm{I},d}\,\frac{V^{t}_{k}}{V_{k}}
 +\mathrm{a}_{\mathrm{P},d}
\label{eq:zip_factor}
\end{equation}
with $\mathrm{b}^{t}_{d(k)}$ defined analogously from $(\mathrm{b_{Z}},\mathrm{b_{I}},\mathrm{b_{P}})$. The ZIP model makes the transient power of each load dependent on the bus
voltage $V^{t}_{k}$, with the dependence determined by its load-specific ZIP coefficients. Consequently, loads with different transient behaviors contribute differently to the transient dynamics, even when connected to the same bus. The ZIP model is adopted here for its widespread use. Other differentiable load models, either static or dynamic, can replace \eqref{eq:zip_factor}, and $\mathrm{P}_{d(k)}$ then enters the analysis of Section~\ref{sec:economics_tscopf} through the derivative of the adopted model with respect to it.

\subsection{Transient Stability Constraints}
\label{subsec:stability_constraints}

Transient stability is the ability of a power system to keep its generators in synchronism after a large disturbance~\cite{Kundur1994, IEEE_TF_2021}, and is commonly assessed from the relative motion of the generator angles. As in previous works, the \ac{COI}, defined as the inertia-weighted average angle, is taken as the common reference~\cite{zarateminano2010, liederer2022, xingrui_2026_pce}. The \ac{TS} criterion bounds the angular deviation of each generator with respect to the \ac{COI},
\begin{subequations}
\label{eq:angle_wrt_COI}
\begin{align}
    (\overline{\Gamma}^{\,t}_{g})\quad
    & \left(\delta_g^t
    - \frac{\sum_{i\in\mathcal{G}} \mathrm{H}_i \delta_i^t}
    {\sum_{i\in\mathcal{G}} \mathrm{H}_i}\right)
    - \updelta_g^{\max}
    \leq 0 & \forall \, t \in \mathcal{T},\; g\in\mathcal{G}
    \label{eq:ineq_angle_coi_max}\\
    (\underline{\Gamma}^{\,t}_{g})\quad
    & \updelta_g^{\min}
    - \left(\delta_g^t
    - \frac{\sum_{i\in\mathcal{G}} \mathrm{H}_i \delta_i^t}
    {\sum_{i\in\mathcal{G}} \mathrm{H}_i}\right)
    \leq 0 & \forall \, t\in\mathcal{T},\; g\in\mathcal{G}
    \label{eq:ineq_angle_coi_min}
\end{align}
\end{subequations}
with $\mathrm{H}_i$ the inertia constant of generator $i$. The bounds are usually symmetric, $\updelta_g^{\min}=-\updelta_g^{\max}$, and $\pm100^\circ$ is a common empirical threshold~\cite{liederer2022, gan2000, zarateminano2010}. This value is a proxy, since the actual stability boundary depends on the system, the operating point and the disturbance. Alternative criteria based on rotor speed, on combined angle and speed deviations, or on pairwise rotor-angle differences~\cite{aghahassani2024, arredondo2023, wu_2024_nn_uctscopf} may replace \eqref{eq:angle_wrt_COI} without altering the analysis of Section~\ref{sec:economics_tscopf}, provided they remain differentiable in the trajectory variables.

\subsection{Complete Optimization Model for TSCOPF}
\label{subsec:tscopf_complete}

Collecting the steady-state \ac{ACOPF}, the discretized \acp{DAE}, the transient network equations and the \ac{TS} constraints gives the complete \ac{TSCOPF} model,
\begin{subequations}
\label{eq:tscopf}
\begin{align}
& \hspace*{-14mm} \min\;\;
  \sum_{g \in \mathcal{G}} \mathrm{c}_g P_g
\label{eq:obj_tscopf}
&&\\[-1pt]
& \hspace*{-13.5mm} \text{subject to:}\quad
  \eqref{eq:pbalance}-\eqref{eq:stlim},
  \eqref{eq:dynamic_constraints},
  \eqref{eq:net_dyn},
  \eqref{eq:angle_wrt_COI}
&&
\end{align}
\end{subequations}

For clarity, although \eqref{eq:tscopf} is written for a single contingency, a set of critical contingencies, typically identified beforehand by the system operator, can be handled by replicating \eqref{eq:dynamic_constraints}, \eqref{eq:net_dyn}, and \eqref{eq:angle_wrt_COI} for each contingency, all sharing the steady-state variables $\boldsymbol{\mathrm{x}}$. Any disturbance that can be represented through the network admittances or the model parameters during the transient, e.g., short circuits, branch outages, generator and load disconnections, fits this structure.

\section{Shadow Prices and Settlement in TSCOPF}
\label{sec:economics_tscopf}

The \ac{TSCOPF} of Section~\ref{sec:tscopf} yields shadow prices that differ from those of \ac{ACOPF}. This section first establishes the conventional \ac{LMP} interpretation of \ac{ACOPF} as a reference and then derives the shadow prices associated with loads and generators in \ac{TSCOPF}. Section~\ref{subsec:stab_sensitivity} relates their stability components to the sensitivity of the \ac{TS} constraints \eqref{eq:angle_wrt_COI}, and Section~\ref{subsec:settlement} defines the resulting settlement.

\subsection{Locational Marginal Prices in an ACOPF}
\label{subsec:LMP_ACOPF}

Consider the Lagrangian function of the steady-state \ac{ACOPF} \eqref{eq:acopf}:
\begin{equation}
\begin{aligned}
    & L_{\mathrm{acopf}}
    =
    \sum_{g \in \mathcal{G}} \mathrm{c}_g P_g +
    \sum_{k \in \mathcal{B}}\lambda_k\Bigl(P_k + \sum_{d\in\mathcal{D}_k}\mathrm{P}_{d(k)}- \sum_{g\in\mathcal{G}_k}P_{g(k)}\Bigr)
    \\
    & 
    + \sum_{g \in \mathcal{G}}\overline{\mu}_g\left(P_g- \mathrm{P}^{\max}_g\right) + \sum_{g \in \mathcal{G}}\underline{\mu}_g\left(\mathrm{P}^{\min}_g - P_g\right) + L_{r-\mathrm{acopf}} \\
\end{aligned}
\label{eq:lagrangian_acopf}
\end{equation}
where $L_{r-\mathrm{acopf}}$ collects the remaining terms associated to constraints \eqref{eq:qbalance}, \eqref{eq:ref}, and \eqref{eq:qgmax}-\eqref{eq:stlim}.

Litvinov~\cite{litvinov2010_lmp} defines the \ac{LMP} as `\emph{a change in production cost to optimally deliver an increment of load at the location, while satisfying all the constraints}'. For the sake of simplicity and without loss of generality, demand at each bus is considered to be fixed and inelastic. Under this assumption, by the Envelope Theorem \cite{carter_2001_envelope}, the \ac{LMP} at bus $k$ can be obtained as the partial derivative of the Lagrangian with respect to the bus demand, evaluated at the optimal solution~\cite{chen2002, litvinov2010_lmp, WANG2020_lmp}:
\begin{equation}
    \frac{\partial L_{\mathrm{acopf}}}{\partial \mathrm{P}_{d(k)} } = \lambda_{k} \; \left[\frac{\$}{\mathrm{MWh}}\right]
    \label{eq:lmp_opf_def}
\end{equation}

Under the sign convention of \eqref{eq:lagrangian_acopf}, this sensitivity is given by the dual variable $\lambda_k$, which reflects the marginal effect of an incremental demand at bus $k$ on the optimal production cost, including the effects of losses and network constraints. It is a bus-level quantity: every load at bus $k$ enters \eqref{eq:pbalance} with the same coefficient, so it is charged the nodal energy price $\lambda_k$, regardless of its voltage sensitivity.

Let $P_{g(k)}$ denote the active power of generator $g$ at bus $k$. The stationarity condition with respect to $P_{g(k)}$ derived from \eqref{eq:lagrangian_acopf} relates the nodal energy price $\lambda_k$ to the generator's marginal operating cost and the dual variables of its output limits:
\begin{align}
    \frac{\partial L_{\mathrm{acopf}}}{\partial P_{g(k)} }
    & = \mathrm{c}_g - \lambda_{k} + \overline{\mu}_g - \underline{\mu}_g = 0 \notag\\
    \lambda_{k}
    & = \mathrm{c}_g + \overline{\mu}_g - \underline{\mu}_g \quad \forall \; g \in \mathcal{G}_k
    \label{eq:stat_pg_acopf}
\end{align}

Complementary slackness shows that the relationship between the nodal energy price and a generator's marginal production cost depends on its operating limits. For a generator operating strictly within its limits, $\lambda_k=\mathrm{c}_g$. When the upper generation limit is binding, $\lambda_k=\mathrm{c}_g+\overline{\mu}_g\geq \mathrm{c}_g$, whereas when the lower generation limit is binding, $\lambda_k=\mathrm{c}_g-\underline{\mu}_g\leq \mathrm{c}_g$.  Thus, in \ac{ACOPF}, any difference between the nodal energy price and the marginal cost of a generator stems solely from the dual variables of its output limits.

\subsection{Shadow Price Associated with Loads in \ac{TSCOPF}}
\label{subsec:price_load}

In TSCOPF, a load enters both the steady-state power balance \eqref{eq:pbalance} and the transient power balances \eqref{eq:net_p_ts} at every time step. Therefore, its associated shadow price includes contributions from both the steady-state and transient constraints. The former yields a nodal energy component, whereas the latter gives rise to a stability component that depends on the load location, its dynamic behaviour, and the disturbance considered.

The Lagrangian of the \ac{TSCOPF} extends \eqref{eq:lagrangian_acopf} with the transient constraints:
\begin{equation}
\begin{aligned}
& L=\; L_{\mathrm{acopf}} + \sum_{g\in\mathcal{G}}\left(\boldsymbol{\tau}^{0\top}_{g}\boldsymbol{h}^{0}_{\tau,g} \right) 
 + \sum_{t\in\mathcal{T}}\sum_{g\in\mathcal{G}}
 \left(\boldsymbol{\tau}^{t\top}_{g}\boldsymbol{h}^{t}_{\tau,g} \right) \\
& + \sum_{t\in\mathcal{T}}\sum_{k\in\mathcal{B}}
 \left(\hat{\lambda}^{t}_{k}\,\hat{h}^{t}_{P,k} + \hat{\beta}^{t}_{k} \, \hat{h}^{t}_{Q,k}\right) + \sum_{t\in\mathcal{T}}\sum_{g\in\mathcal{G}}
 \left(\overline{\Gamma}^{\,t}_{g}\,\overline{s}^{\,t}_{g}
      +\underline{\Gamma}^{\,t}_{g}\,\underline{s}^{\,t}_{g}\right)
\end{aligned}
\label{eq:lagrangian_tscopf}
\end{equation}
where $\boldsymbol{h}^{0}_{\tau,g}$ and $\boldsymbol{h}^{t}_{\tau,g}$ denote the discretized \acp{DAE}
\eqref{eq:dynamic_constraints}, $\hat{h}^{t}_{P,k}$ and $\hat{h}^{t}_{Q,k}$ the transient network equations \eqref{eq:net_dyn},  and $\overline{s}^{t}_{g}$ and $\underline{s}^{t}_{g}$ the stability constraints \eqref{eq:angle_wrt_COI} of Section~\ref{sec:tscopf}.

With the ZIP factor $\mathrm{a}^{t}_{d(k)}$ of \eqref{eq:zip_factor}, the Envelope Theorem applied to \eqref{eq:lagrangian_tscopf} gives
\begin{equation}
    \frac{\partial L}{\partial \mathrm{P}_{d(k)}}
    = \lambda_{k}
    + \sum_{t\in\mathcal{T}}\hat{\lambda}^{t}_{k}\,\mathrm{a}^{t}_{d(k)}
    \; \left[\frac{\$}{\mathrm{MWh}}\right]
    \label{eq:dL_dPd_tscopf}
\end{equation}

The factor $\mathrm{a}^{t}_{d(k)}$ is evaluated at the optimal voltage at each time step, retaining the voltage dependence of the load along the optimal transient trajectory. Hence, the increment in production cost associated with optimally serving an increment of load $d$ at bus $k$ is denoted by the shadow price $\pi_{d(k)}$, such that
\begin{subequations}
\begin{align}
    \pi_{d(k)} &= \lambda_k + \psi_{d(k)}
    \label{eq:pi_load}\\
    \psi_{d(k)} &= \sum_{t\in\mathcal{T}} \hat{\lambda}^{t}_{k}\,\mathrm{a}^{t}_{d(k)}
    \label{eq:psi_load}
\end{align}
\end{subequations}
Here, $\lambda_k$ retains the nodal energy component associated with the steady-state power balance constraint \eqref{eq:pbalance}, while $\psi_{d(k)}$ is a load-specific stability component arising from the transient dynamics through \eqref{eq:net_p_ts}. Thus, co-located loads at the same bus are subject to the same nodal energy component $\lambda_k$, but may have different stability components $\psi_{d(k)}$ if they exhibit different transient behaviors. Since the stability component $\psi_{d(k)}$ is unrestricted in sign, it can either increase or decrease the shadow price $\pi_{d(k)}$ relative to $\lambda_k$. If $\psi_{d(k)}$ is sufficiently negative that $\pi_{d(k)}<0$, the load is paid $|\pi_{d(k)}|$.

\subsection{Shadow Price Associated with Generators in \ac{TSCOPF}}
\label{subsec:price_gen}

Transient stability is governed by the power imbalance that a disturbance creates at each generator, that is, the difference between its input power and the electrical power it delivers to the network. This imbalance drives the generator angles apart, and stability is retained when the angle differences remain bounded. For \acp{SG}, the imbalance is the difference between mechanical and electrical power and the resulting dynamics are described by the swing equation \cite{Kundur1994}, while for \acp{IBG} they are determined by their control laws. In both cases, the active-power dispatch $P_g$ fixes the pre-fault operating point, which sets the initial conditions of the transient dynamics when the disturbance occurs. Although system inertia also strongly influences \ac{TS}, it is held fixed in \ac{TSCOPF} because commitment is determined beforehand and no commitment variables are included. Among the decision variables of the \ac{TSCOPF}, $P_g$ is the only one that carries a production cost and is remunerated in the market. The shadow price of the \ac{TS} constraints is therefore derived with respect to $P_g$. 

In \eqref{eq:tscopf}, $P_g$ appears in the objective function, in the pre-fault balance \eqref{eq:pbalance}, in the output limits \eqref{eq:pgmax}--\eqref{eq:pgmin}, and in the initialization constraints \eqref{eq:init_Pe}--\eqref{eq:init_Pm}. It does not appear explicitly in any other constraint. Hence, the stationarity condition with respect to $P_{g(k)}$ derived from \eqref{eq:lagrangian_tscopf} is
\begin{align}
\frac{\partial L}{\partial P_{g(k)}} & =
\frac{\partial L_{\mathrm{acopf}}}{\partial P_{g(k)}} + \frac{\partial \left(
\sum_{i \in \mathcal{G}}
\boldsymbol{\tau}^{0\top}_{i}\boldsymbol{h}^{0}_{\tau,i}
\right)
}{\partial P_{g(k)}} \notag
\\
& =
\mathrm{c}_g
-\lambda_{k}
+\overline{\mu}_g
-\underline{\mu}_g
+\sigma^{0}_{\mathrm{P},g}
+\chi^{0}_{\mathrm{m},g}
=0
\quad \forall g \in \mathcal{G}_k
\label{eq:stat_pg}
\end{align}
\noindent and the nodal energy component at bus $k$ must satisfy
\begin{equation}
\lambda_{k}
=
\mathrm{c}_g
+\overline{\mu}_g
-\underline{\mu}_g
+\sigma^{0}_{\mathrm{P},g}
+\chi^{0}_{\mathrm{m},g}
\quad \forall g \in \mathcal{G}_k
\label{eq:lambda_gen_tscopf}
\end{equation}

The first three terms on the right-hand side of \eqref{eq:lambda_gen_tscopf} have the same interpretation as in the \ac{ACOPF} relation \eqref{eq:stat_pg_acopf}. The additional terms $\sigma^{0}_{\mathrm{P},g}$ and $\chi^{0}_{\mathrm{m},g}$ arise from the equality constraints that couple $P_g$ to the initialization of the dynamic model \eqref{eq:initial_coupling}, and therefore quantify the effect of the generator's dispatch on the \ac{TS} constraints.

Although $\sigma^{0}_{\mathrm{P},g}$ and $\chi^{0}_{\mathrm{m},g}$ are not associated with the generator's output limits \eqref{eq:pgmax}--\eqref{eq:pgmin}, they act in \eqref{eq:lambda_gen_tscopf} analogously to the dual variables of these limits. To illustrate this, consider a generator operating within its output limits in the \ac{TSCOPF} solution, for which $\overline{\mu}_g=\underline{\mu}_g=0$. If an increase in its output pushes the system toward the stability limit, the \ac{TS} constraints hold its output below the level that would minimize the production cost, even though its upper limit $\mathrm{P}^{\max}_{g}$ is not reached. In this case, $\sigma^{0}_{\mathrm{P},g}+\chi^{0}_{\mathrm{m},g}>0$ and $\lambda_k>\mathrm{c}_g$, as for a generator at its upper limit. Conversely, if a decrease in its output pushes the system toward the stability limit, the \ac{TS} constraints hold its output above the level that would minimize the production cost, even though its lower limit $\mathrm{P}^{\min}_{g}$ is not reached. In this case, $\sigma^{0}_{\mathrm{P},g}+\chi^{0}_{\mathrm{m},g}<0$ and $\lambda_k<\mathrm{c}_g$, as for a generator at its lower limit, that is, the generator does not recover its operating cost when paid $\lambda_k$ alone.

However, two differences with respect to an output limit arise. First, the restriction stems from the system-wide stability effect and not from the actual capability limit of a single generator. Second, although $\overline{\mu}_g$ and $\underline{\mu}_g$ are also generator-specific, they concern only the capability of the generator itself, and an increment of power at bus $k$ has the same effect on the rest of the system regardless of which generator supplies it. In contrast, $\sigma^{0}_{\mathrm{P},g}+\chi^{0}_{\mathrm{m},g}$ measures the effect of the generator's output on the \ac{TS} constraints, which are shared by the whole system, and this effect depends on the dynamic characteristics of each unit. An increment of power at bus $k$ therefore affects the system differently depending on the generator that supplies it. Since the nodal energy price $\lambda_k$ values an increment of power only in the pre-fault balance \eqref{eq:pbalance} and is common to all assets at bus $k$, it cannot reflect the individual contribution of each generator to maintaining stability. This motivates the definition of a generator-specific shadow price:
\begin{subequations}
\begin{align}
    \pi_{g(k)} &= \lambda_k + \psi_{g(k)} \quad \forall g \in \mathcal{G}_k
    \label{eq:pi_gen}\\
    \psi_{g(k)} &= -\left(\sigma^{0}_{\mathrm{P},g}+\chi^{0}_{\mathrm{m},g}\right) \quad \forall g \in \mathcal{G}_k
    \label{eq:psi_gen}
\end{align}
\end{subequations}
where $\psi_{g(k)}$ is the generator-specific stability component, and $\pi_{g(k)}$ is the shadow price used for settlement. Essentially, the shadow price $\pi_{g(k)}$ values an increment of power from generator $g$ through both its contribution to the pre-fault balance \eqref{eq:pbalance} and its effect, through the initialization \eqref{eq:initial_coupling}, on the \ac{TS} constraints, as \eqref{eq:dL_dPd_tscopf} does for a load. Substituting \eqref{eq:lambda_gen_tscopf} into \eqref{eq:pi_gen} gives
\begin{equation}
    \pi_{g(k)} = \mathrm{c}_g + \overline{\mu}_g - \underline{\mu}_g \quad \forall g \in \mathcal{G}_k
    \label{eq:pi_gen_reduced}
\end{equation}
which reproduces the \ac{ACOPF} relation \eqref{eq:stat_pg_acopf} with $\pi_{g(k)}$ in place of $\lambda_k$. 

As with the loads, co-located generators are paid the same nodal energy component $\lambda_k$, but may have different stability components $\psi_{g(k)}$ according to their transient behaviors. Since $\psi_{g(k)}$ is unrestricted in sign, the stability component can either increase or decrease the shadow price $\pi_{g(k)}$ relative to the nodal energy component $\lambda_k$.

Garcia \emph{et al.}~\cite{garcia2025} obtained an expression analogous to \eqref{eq:pi_gen}, with the stability term given by the gradient with respect to $P_g$ of a neural-network surrogate representing the stability constraint. Modeling generators as price-taking profit maximizers, they showed that this price supports the cost-minimizing dispatch. The same property follows here from \eqref{eq:pi_gen_reduced}, since $P_g$ maximizes the profit `$(\pi_{g(k)}-\mathrm{c}_g)P_g$' of generator $g$ within its output limits. In \cite{garcia2025}, this price remains an analytical expression: it is neither evaluated in the case studies nor used for settlement, since the focus of that work is the accuracy of the surrogate. In this paper, $\psi_{g(k)}$ is obtained from the dual variables of the discretized \acp{DAE} embedded in \ac{TSCOPF}, it is computed for each generator in Section~\ref{sec:case_studies}, and $\pi_{g(k)}$ is used for settlement together with the shadow price associated with loads in Section~\ref{subsec:price_load}, as described in Section~\ref{subsec:settlement}.

\subsection{Sensitivities of the Stability Constraints to Power Demand and Generation}
\label{subsec:stab_sensitivity}

To relate the stability components of the shadow prices associated with loads and generators to the active stability constraints, one may collect the discretized \acp{DAE} \eqref{eq:dynamic_constraints} and the transient network constraints \eqref{eq:net_dyn} into $F(\boldsymbol{z},\boldsymbol{w})=\boldsymbol{0}$, where $\boldsymbol{z}$ contains the dynamic and algebraic  variables and $\boldsymbol{w}$ denotes a vector of active powers of generators and loads. Let $\boldsymbol{s}(\boldsymbol{z})\leq\boldsymbol{0}$ denote the \ac{TS} constraints in \eqref{eq:angle_wrt_COI}, with associated multipliers $\boldsymbol{\Gamma}\geq\boldsymbol{0}$. The Lagrangian function in \eqref{eq:lagrangian_tscopf} can be rewritten as
\begin{equation}
L
=
L_{\mathrm{acopf}}
+
\boldsymbol{\varphi}^{\top}F(\boldsymbol{z},\boldsymbol{w})
+
\boldsymbol{\Gamma}^{\top}\boldsymbol{s}(\boldsymbol{z})
\label{eq:lagr_compact}
\end{equation}
where $\boldsymbol{\varphi}$ contains the multipliers of $F$ and $L_{\mathrm{acopf}}$ is independent of $\boldsymbol{z}$. The relevant Jacobian matrices evaluated at the \ac{KKT} point are
\begin{equation}
J_z=\frac{\partial F}{\partial\boldsymbol{z}},
\qquad
J_w=\frac{\partial F}{\partial\boldsymbol{w}},
\qquad
J_s=\frac{\partial\boldsymbol{s}}{\partial\boldsymbol{z}}
\label{eq:jacobians}
\end{equation}

If $F$ is continuously differentiable and $J_z$ is nonsingular, the implicit function theorem~\cite{krantz2012} guarantees a locally differentiable trajectory $\boldsymbol{z}(\boldsymbol{w})$ satisfying $F(\boldsymbol{z}(\boldsymbol{w}),\boldsymbol{w})=\boldsymbol{0}$, with
\begin{equation}
\frac{d\boldsymbol{z}}{d\boldsymbol{w}}
=
-J_z^{-1}J_w
\label{eq:dz_dw}
\end{equation}

Although $\boldsymbol{s}$ has no explicit dependence on $\boldsymbol{w}$, i.e., \eqref{eq:angle_wrt_COI} contains neither $\mathrm{P}_{d(k)}$ nor $P_{g(k)}$, its value depends on $\boldsymbol{w}$ through the trajectory $\boldsymbol{z}(\boldsymbol{w})$. Hence, its total sensitivity to $\boldsymbol{w}$ is
\begin{equation}
\frac{d\boldsymbol{s}}{d\boldsymbol{w}}
=
\frac{\partial\boldsymbol{s}}{\partial\boldsymbol{z}}
\frac{d\boldsymbol{z}}{d\boldsymbol{w}}
=
-J_sJ_z^{-1}J_w
\label{eq:ds_dw}
\end{equation}

Stationarity of \eqref{eq:lagr_compact} with respect to $\boldsymbol{z}$ gives
\begin{equation}
\boldsymbol{\Gamma}^{\top}J_s = - \boldsymbol{\varphi}^{\top}J_z
\label{eq:gamma_Js}
\end{equation}

Multiplying \eqref{eq:ds_dw} by $\boldsymbol{\Gamma}^{\top}$ and using the relation defined by \eqref{eq:gamma_Js} yields
\begin{equation}
\begin{aligned}
\boldsymbol{\Gamma}^{\top} \frac{d\boldsymbol{s}}{d\boldsymbol{w}}
&=
-\boldsymbol{\Gamma}^{\top}
J_sJ_z^{-1}J_w
\\
&=
\boldsymbol{\varphi}^{\top}
J_zJ_z^{-1}J_w = \boldsymbol{\varphi}^{\top}J_w = \boldsymbol{\varphi}^{\top} \frac{\partial F}{\partial\boldsymbol{w}}
\end{aligned}
\label{eq:sens_identity}
\end{equation}

Eq. \eqref{eq:sens_identity} shows that the multiplier-weighted sensitivity of the \ac{TS} constraints equals the dual contribution of the equalities through which $\boldsymbol{w}$ enters the transient model. When several contingencies are enforced, $F$ and $\boldsymbol{s}$ stack the constraints of all of them and \eqref{eq:sens_identity} holds unchanged. Each asset then has a single stability component, given by the sum of the contributions of the individual contingencies, and a contingency with no active \ac{TS} constraint contributes zero.

For loads, $\mathrm{P}_{d(k)}$ enters $F$ through \eqref{eq:net_p_ts}, whose derivative with respect to $\mathrm{P}_{d(k)}$ is the ZIP factor $\mathrm{a}^{t}_{d(k)}$ defined in \eqref{eq:zip_factor}. Hence,
\begin{equation}
\boldsymbol{\Gamma}^{\top}\frac{d\boldsymbol{s}}{d\mathrm{P}_{d(k)}}
=
\sum_{t\in\mathcal{T}}
\hat{\lambda}_k^t\,\mathrm{a}_{d(k)}^{t}
=
\psi_{d(k)}
\label{eq:sens_load}
\end{equation}
where \eqref{eq:psi_load} has been used.

Similarly, generator's active power $P_{g(k)}$ enters $F$ only through \eqref{eq:init_Pe} and \eqref{eq:init_Pm}. Therefore, for each generator,
\begin{equation}
\boldsymbol{\Gamma}^{\top}\frac{d\boldsymbol{s}}{dP_{g(k)}}
=
\sigma^{0}_{\mathrm{P},g}
+
\chi^{0}_{\mathrm{m},g}
=
-\psi_{g(k)}
\label{eq:sens_gen}
\end{equation}
where \eqref{eq:psi_gen} has been used with opposite sign. 

Notably, to the best of the authors' knowledge, the derivation of these sensitivities has not been reported for a \ac{TSCOPF} with embedded \acp{DAE}. It gives the stability component of generators and loads even though the \ac{TS} constraints contain neither the generated nor the demanded power explicitly.

Since $\boldsymbol{\Gamma}\geq\boldsymbol{0}$, each active stability constraint enters \eqref{eq:sens_identity} with the sign of its own sensitivity, so the product gives the net effect of an asset on the active constraints. For an infinitesimal increase in the power of an asset, a positive component of $\frac{d\boldsymbol{s}}{d\boldsymbol{w}}$ indicates movement of the corresponding constraint function toward its limit, i.e., a reduction of the stability margin, whereas a negative component indicates movement away from it. For a load, \eqref{eq:sens_load} carries this sign into $\psi_{d(k)}$. Hence, a load that reduces the stability margin has $\psi_{d(k)}>0$ and pays more than $\lambda_k$, whereas a load that improves it has $\psi_{d(k)}<0$ and pays less. For a generator, the minus sign in \eqref{eq:sens_gen} reverses this mapping. Thus, a generator that reduces the stability margin has $\psi_{g(k)}<0$ and is paid less than $\lambda_k$, whereas a generator that improves it has $\psi_{g(k)}>0$ and is paid more, in agreement with the two cases of Section~\ref{subsec:price_gen}. 

The stability component therefore charges the assets that reduce the stability margin and remunerates those that improve it. In this sense, the shadow prices allocate the cost of securing \ac{TS} among the assets without a separate allocation rule, since each stability component follows directly from the dual variables of \ac{TSCOPF}.
\subsection{\texorpdfstring{Settlement at the Shadow Prices of TSCOPF}{Settlement at the Shadow Prices of TSCOPF}}
\label{subsec:settlement}

When every asset is settled at its associated shadow price, each load pays $\pi_{d(k)}$ and each generator receives $\pi_{g(k)}$. The settlement residual is the difference between the load payments and the generator revenues:
\begin{equation}
R
=
\sum_{d\in\mathcal{D}}\pi_{d(k)}\,\mathrm{P}_{d(k)}
\;-\;
\sum_{g\in\mathcal{G}}\pi_{g(k)}\,P_{g(k)}
\label{eq:residual}
\end{equation}

Since $\lambda_k$ includes the effects of losses and network congestion, as in \ac{ACOPF}, the shadow prices $\pi_{d(k)}$ and $\pi_{g(k)}$ internalize these effects together with the stability components $\psi_{d(k)}$ and $\psi_{g(k)}$. Hence, $R$ collects in a single quantity the contributions of losses, congestion, and \ac{TS}. For instance, in a lossless and uncongested network, a nonzero residual stems solely from the price spread caused by the stability constraints. If no stability constraint is active, $\boldsymbol{\Gamma}=\boldsymbol{0}$, the stability components vanish, and $R$ reduces to the residual of \ac{ACOPF}. Section~\ref{sec:case_studies} analyzes this settlement in different scenarios.

\section{Case Studies}
\label{sec:case_studies}

The case studies analyze the shadow prices and the settlement that result from the \ac{TS} constraints. The WECC 9-bus system, made lossless and uncongested, is used for isolating the stability component of the shadow prices, first with one generator and one load per bus and then with co-located assets. A modified IEEE 39-bus system then shows that the same properties hold in a network with losses, congestion and \acp{IBG}. The models are implemented in \texttt{Julia} with \texttt{JuMP}~\cite{JuMP2023} and solved with \texttt{Ipopt}, on a laptop with an Intel Core i7-1255U CPU @1.70~GHz and 16~GB of RAM. The source code and related documentation are available on GitHub~\cite{Code}.

\subsection{WECC 9-bus system}
\label{subsec:9bus_system}

\begin{figure}[tb!]
\centering
\includegraphics[width=0.85\columnwidth]{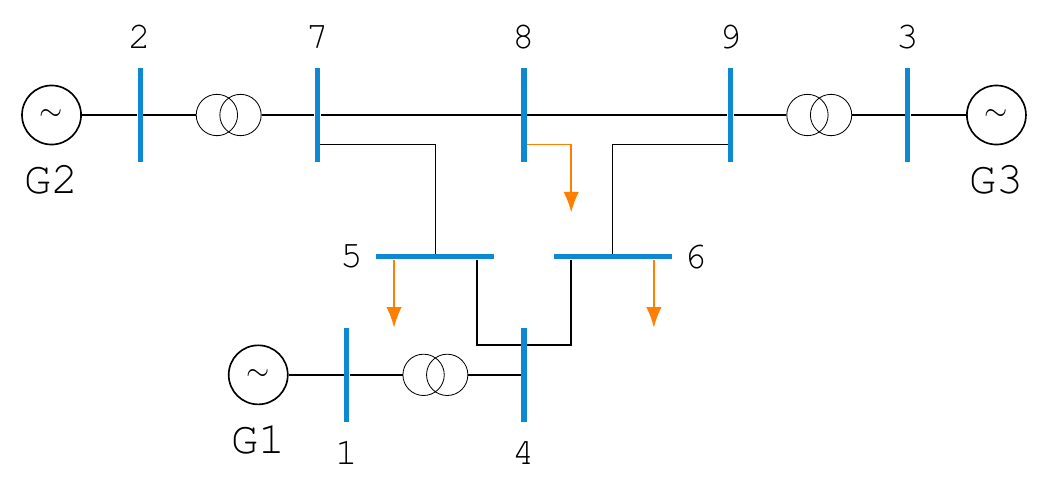}
\caption{Single-line diagram of the WECC 9-bus system.}
\label{fig:wecc9b3m}
\end{figure}

Fig.~\ref{fig:wecc9b3m} shows the WECC 9-bus system. The marginal costs of the three units are $\{\mathrm{c}_{g1},\mathrm{c}_{g2},\mathrm{c}_{g3}\}=\{50,30,10\}\;\$/\mathrm{MWh}$. Network parameters are taken from the MATPOWER database~\cite{MATPOWER}, and the dynamic data of the generators, exciters and turbine governors are given in~\cite{Code}. The original load is scaled by a factor of 1.5 to obtain a more demanding operating condition.

\subsubsection{Shadow Prices in a Lossless and Uncongested Network}
\label{subsubsec:prices_9bus}

This analysis isolates the effect of \ac{TS} constraints on the shadow prices. To this end, the network is made lossless, and uncongested by removing the branch-flow limits \eqref{eq:sflim}--\eqref{eq:stlim}. Under these conditions, any price difference observed in \ac{TSCOPF} stems solely from the \ac{TS} constraints, as discussed in Section~\ref{subsec:settlement}. \ac{ACOPF} is solved first and serves as the reference. \ac{TSCOPF} then adds the \ac{TS} constraints for a single contingency, a three-phase short circuit at bus 7 cleared after $100\;\mathrm{ms}$ by opening branch 5--7. This fault duration, five cycles at the nominal frequency of $50$~Hz, accounts for relay detection and circuit-breaker operation, in line with industry practice. During the transient, all loads are modeled as constant impedances. The \ac{TS} constraints \eqref{eq:angle_wrt_COI} bound the angles within $\pm100^\circ$ of the \ac{COI} over a window of $5\;\mathrm{s}$, discretized with a time step of $5\;\mathrm{ms}$.

The \ac{TS} constraints change the least-cost dispatch substantially, and the total operating cost increases from $8.78$~k\$/h in \ac{ACOPF} to $11.67$~k\$/h in \ac{TSCOPF}, i.e., by $32.9\%$. In \ac{ACOPF}, G3 operates at its upper limit of $270$~MW, G2 supplies the remaining $202.5$~MW, and G1, the most expensive unit, produces no active power. In \ac{TSCOPF}, the outputs of G2 and G3 are reduced to $89$ and $254.6$~MW, respectively, and G1 is dispatched at $128.9$~MW. No generator operates at its output limits in \ac{TSCOPF}, i.e., $\overline{\mu}_g=\underline{\mu}_g=0$ for the three units, which makes this case a direct illustration of \eqref{eq:lambda_gen_tscopf}.

\begin{figure}[!tb]
    \centering
    \includegraphics[width=0.95\columnwidth]{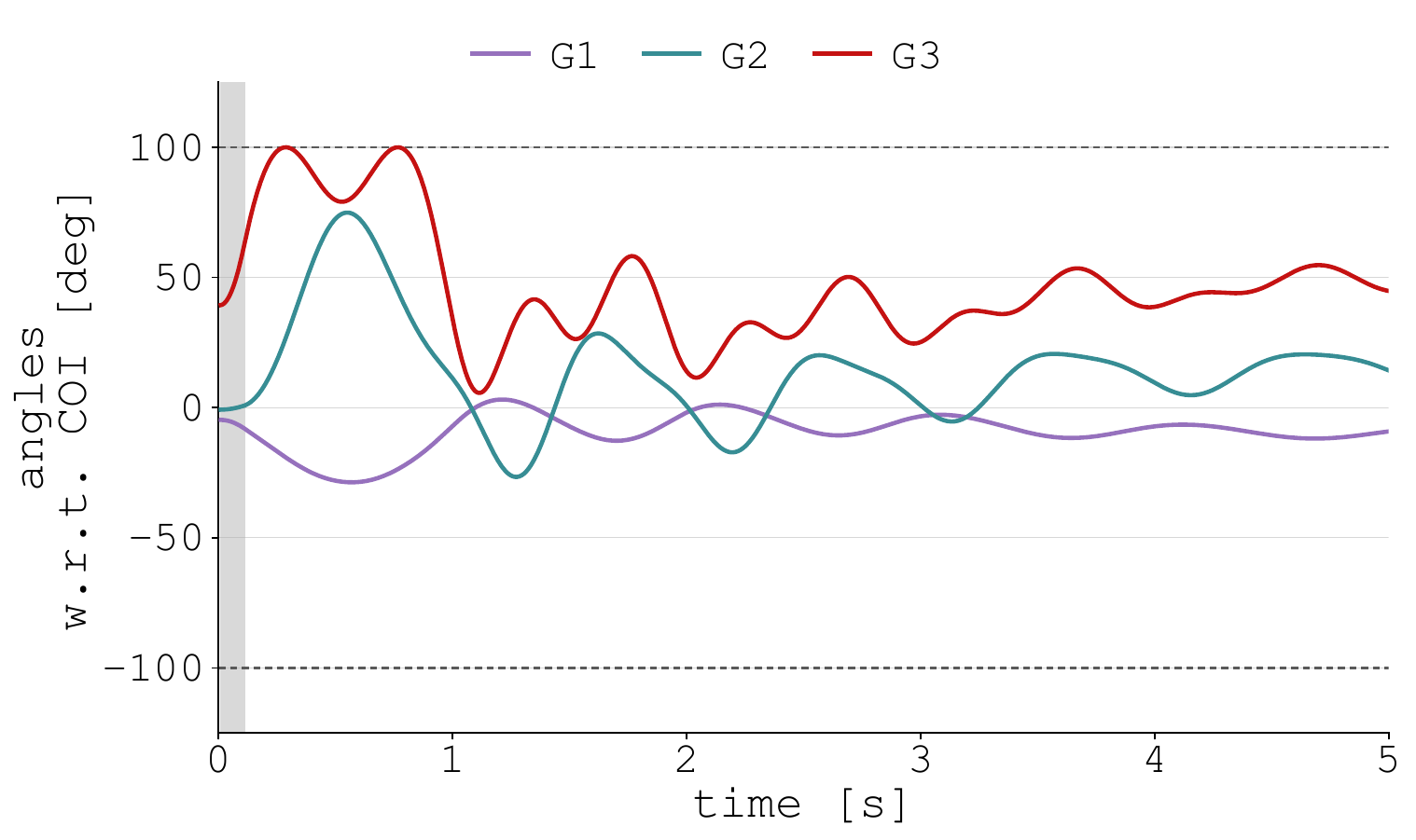}
    \caption{Angles relative to the \ac{COI} of the three generators obtained with \ac{TSCOPF} for the short circuit at bus 7. The shaded area represents the short circuit before being cleared.}
    \label{fig:angles_9bus}
\end{figure}

Fig.~\ref{fig:angles_9bus} shows the angles of the three generators relative to the \ac{COI} obtained with \ac{TSCOPF}. The upper stability limit \eqref{eq:ineq_angle_coi_max} is active only for G3, the unit with the lowest inertia. During the short circuit, the acceleration of G3 increases with its pre-fault output, which explains the reduction of its output compared to \ac{ACOPF}. In turn, the \ac{COI} essentially follows the rotor of G1, which has the highest inertia among the three units. A higher output of G1 therefore increases the acceleration of the \ac{COI} during the fault, reducing the angular deviation of G3 relative to the \ac{COI}. This increase is balanced by the lower outputs of G2 and G3.

The sign of the stability component identifies how each generator affects the stability margin, while the nodal energy component alone does not reflect it. \ac{ACOPF} yields a nodal energy price of $30.00$~\$/MWh at every bus, set by the marginal generator G2. In turn, Table~\ref{tab:prices_9bus} shows that the nodal energy component $\lambda_k$ of the shadow prices in \ac{TSCOPF} ranges from $40.37$ to $48.07$~\$/MWh at the load and generator buses, although the network is lossless and uncongested. Specifically, $\lambda_k$ lies $8.86$~\$/MWh below the marginal cost of G1, and $12.55$ and $38.07$~\$/MWh above those of G2 and G3, as established in \eqref{eq:lambda_gen_tscopf}. Paid $\lambda_k$ alone, G1 would not recover its operating cost, with a loss of $1.14$~k\$/h, while G3, whose output is restricted by the stability limit, would be paid almost five times its marginal cost. The shadow price $\pi_{g(k)}$ removes both effects, since $\pi_{g(k)}=\mathrm{c}_g$ for the three units, and each of them recovers exactly its operating cost.

\begin{table}[!tb]
\captionsetup{justification=centering, textfont={sc,footnotesize}, labelfont=footnotesize, labelsep=newline}
\centering
\caption{TSCOPF results for the lossless and uncongested WECC 9-bus system}
\label{tab:prices_9bus}
\footnotesize
\setlength{\tabcolsep}{4pt}
\begin{tabular}{lcrrrrr}
\toprule
 & Bus & $P$ & $\lambda_k$ & $\psi$ & $\pi$ & Rev./Payment \\
 &  & [MW] & [\$/MWh] & [\$/MWh] & [\$/MWh] & at $\pi$ [k\$/h] \\
\midrule
\multirow{3}{*}{Gen} & 1 & 128.9 & 41.14 & $+$8.86 & 50 & 6.45 \\
 & 2 & 89 & 42.55 & $-$12.55 & 30 & 2.67 \\
 & 3 & 254.6 & 48.07 & $-$38.07 & 10 & 2.55 \\
\midrule
\multirow{3}{*}{Load} & 5 & 187.5 & 40.37 & $+$4.93 & 45.30 & 8.49 \\
 & 6 & 135 & 40.85 & $+$1.55 & 42.40 & 5.72 \\
 & 8 & 150 & 41.73 & $-$7.44 & 34.29 & 5.14 \\
\bottomrule
\end{tabular}
\end{table}

The same sign rule applies to the loads. The stability components are positive at buses 5 and 6, whose loads pay above the nodal energy component of their buses, and negative at bus 8, whose load pays $7.44$~\$/MWh below it. Since bus 8 is adjacent to the faulted bus, an increment of its constant-impedance demand absorbs part of the power that would otherwise accelerate G3, reducing its angular excursion. A load is therefore charged or remunerated according to its own effect on the active stability constraint, as a generator is.

The stability constraints affect the settlement even in the absence of losses and congestion. In \ac{ACOPF}, the settlement residual \eqref{eq:residual} is zero, since the network is lossless and uncongested. At the shadow prices of \ac{TSCOPF}, the loads pay $19.35$~k\$/h and the generators receive $11.67$~k\$/h, resulting in $R=7.68$~k\$/h. With no losses or congestion, this residual is caused entirely by the \ac{TS} constraints, and its value is specific to the settlement at the shadow prices of Section~\ref{subsec:settlement}. It should be noted that a settlement residual also exists in energy markets that price marginal losses, where its allocation follows the rules adopted by each system operator~\cite{isone_m28, pjm_m28}. In this regard, the allocation of the residual caused by the \ac{TS} constraints warrants further investigation.

\subsubsection{Shadow Prices for Co-located \texorpdfstring{Assets}{Assets}}
\label{subsubsec:disc_prices_9bus}

This analysis shows that assets connected at the same bus share the nodal energy component and are nevertheless subject to different shadow prices, which a price defined per bus cannot reflect. To this end, the lossless and uncongested system of Section~\ref{subsubsec:prices_9bus} is modified in two ways. First, a second generator, G1b, is connected at bus 1 next to G1. G1b represents a cheaper and smaller unit, with a marginal cost of $20$~\$/MWh, a capacity of $100$~MW and the machine data of G3, including its inertia constant of $3$~s. Second, the load at bus 5 is split into two loads of $93.75$~MW, one modeled as a constant impedance and the other as a constant current, i.e., with ZIP coefficients $(1,0,0)$ and $(0,1,0)$, respectively. The disturbance and all remaining settings are those of Section~\ref{subsubsec:prices_9bus}. In \ac{ACOPF}, G1b and G3 operate at their upper limits, G2 supplies the remaining $102.5$~MW, and G1 produces no active power. G2 is again the marginal generator and sets the nodal energy price of $30.00$~\$/MWh at every bus.

Table~\ref{tab:colocated_bus1_9bus} lists the results for every generator and load. G1 and G1b share the nodal energy component of $41.60$~\$/MWh, and both improve the stability margin, with stability components of $8.40$~\$/MWh for G1 and $5.40$~\$/MWh for G1b. Since G1 operates within its output limits, $\pi_{g(k)}=\mathrm{c}_g=50.00$~\$/MWh and it recovers exactly its operating cost. G1b is dispatched at its upper limit, and its associated shadow price of $47.00$~\$/MWh exceeds its marginal cost by $27.00$~\$/MWh, the value of the dual variable of that limit. Of this margin, $21.60$~\$/MWh stem from the nodal energy component, which the \ac{TS} constraints raise above the marginal cost of G1b, and $5.40$~\$/MWh from its stability component. A price defined per bus would pay G1 and G1b alike, whereas their stability components differ because each unit affects the stability margin differently.

\begin{table}[!tb]
\captionsetup{justification=centering, textfont={sc,footnotesize}, labelfont=footnotesize, labelsep=newline}
\centering
\caption{TSCOPF results for the WECC 9-bus system with co-located assets 
(generators are named as in Fig.~\ref{fig:wecc9b3m}, and the subscript of each load denotes its bus)}
\label{tab:colocated_bus1_9bus}
\footnotesize
\setlength{\tabcolsep}{2pt}
\begin{tabular}{lrrrrrrr}
\toprule
 & Asset & $P$ & $\lambda_k$ & $\psi$ & $\pi$ & Rev./Paym. & Profit at $\pi$ \\
 &  & [MW] & [\$/MWh] & [\$/MWh] & [\$/MWh] & at $\pi$ [k\$/h] & [k\$/h] \\
\midrule
\multirow{4}{*}{Gen} & G1 & 35.5 & 41.60 & $+$8.40 & 50 & 1.78 & 0.00 \\
 & G1b & 100 & 41.60 & $+$5.40 & 47 & 4.70 & 2.70 \\
 & G2 & 89.5 & 42.60 & $-$12.60 & 30 & 2.69 & 0.00 \\
 & G3 & 247.5 & 48.20 & $-$38.20 & 10 & 2.48 & 0.00 \\
\midrule
\multirow{4}{*}{Load} & Z$_5$ & 93.75 & 40.60 & $+$4.50 & 45.10 & 4.23 & --- \\
 & I$_5$ & 93.75 & 40.60 & $+$5.70 & 46.30 & 4.34 & --- \\
 & Z$_6$ & 135 & 41.20 & $+$1.00 & 42.20 & 5.70 & --- \\
 & Z$_8$ & 150 & 41.90 & $-$7.40 & 34.50 & 5.18 & --- \\
\bottomrule
\end{tabular}
\end{table}

The same holds on the demand side, where the difference stems from the transient behavior of each load. The two loads at bus 5 share the nodal energy component of $40.60$~\$/MWh, but the stability component of the constant-current load exceeds that of the constant-impedance load. As a result, the constant-current load pays $1.20$~\$/MWh more. This difference follows from \eqref{eq:psi_load}: both loads share the dual variables $\hat{\lambda}^{t}_{5}$, and only the factors $\mathrm{a}^{t}_{d(k)}$ differ. While the stability constraints are active, the voltage at bus 5 remains below its pre-fault value, and the demand of the constant-current load, proportional to $V^{t}_{5}/V_{5}$, exceeds that of the constant-impedance load, proportional to $\left(V^{t}_{5}/V_{5}\right)^{2}$. Additional demand at bus 5, supplied mainly by the generators at bus 1 once branch 5--7 is opened, decelerates these generators and increases the angle separation of G3 with respect to the \ac{COI}. The load that draws more power at this bus during the transient therefore carries the larger stability component, whereas demand at bus 8 absorbs part of the power that accelerates G3 and is remunerated.

At the shadow prices, the loads pay $19.45$~k\$/h and the generators receive $11.65$~k\$/h, resulting in $R=7.80$~k\$/h, close to that of the previous case.

\subsection{Modified IEEE 39-bus system}
\label{subsec:39bus_system}

\begin{figure}[!tb]
    \centering
    \includegraphics[width=0.98\columnwidth]{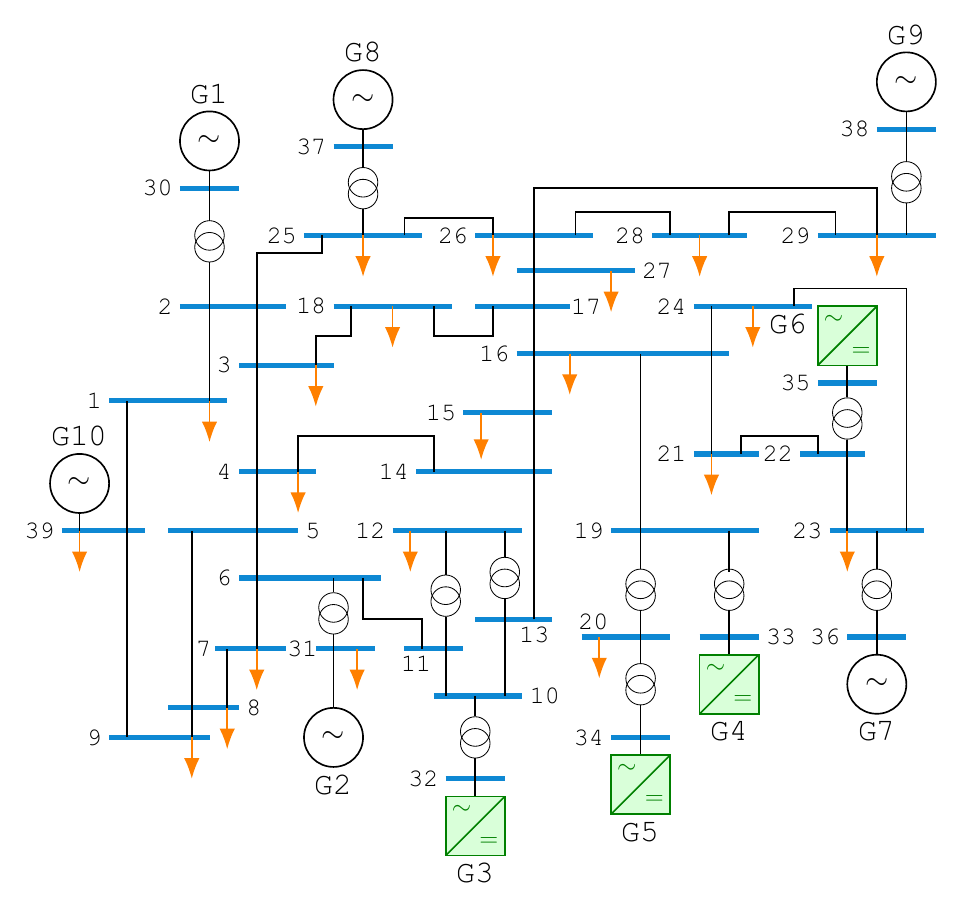}
    \caption{Single-line diagram of the modified IEEE 39-bus system.}
    \label{fig:39bus_system}
\end{figure}

Fig.~\ref{fig:39bus_system} shows the modified IEEE 39-bus system considered in this analysis, whose base data are given in~\cite{babaeinejadsarookolaee_2019_pglib}. It comprises six \acp{SG} and four \acp{IBG}, the latter connected at buses 32 to 35, with a total generation capacity of $7{,}450$~MW for a demand of $5{,}856.4$~MW. The \acp{SG} are represented as in Section~\ref{subsec:9bus_system}. Each \ac{IBG} follows the droop-controlled grid-forming model specified in~\cite{du_2021_gfmdrp}, which includes an active-power droop, a voltage control loop, and a current limiter, here set at $1.2$~pu of the converter rating. The \acp{IBG} enter the \ac{COI} with a virtual equivalent inertia derived from their droop gain and the time constant of their power filter \cite{gfm_in_coi_2025}. In contrast to Section~\ref{subsec:9bus_system}, the network includes losses and the branch-flow limits \eqref{eq:sflim}--\eqref{eq:stlim}. All loads are modeled as constant impedances during the transient. The contingency is a three-phase short circuit at bus 23, cleared after $100\;\mathrm{ms}$ by opening line 23--24. The \ac{TS} constraints \eqref{eq:angle_wrt_COI} bound the angle of the ten units within $\pm100^\circ$ of the \ac{COI} over a window of $5\;\mathrm{s}$, discretized with a time step of $5\;\mathrm{ms}$. \ac{ACOPF} is solved first and serves as the reference.

\begin{figure}[!tb]
    \centering
    \includegraphics[width=0.95\columnwidth]{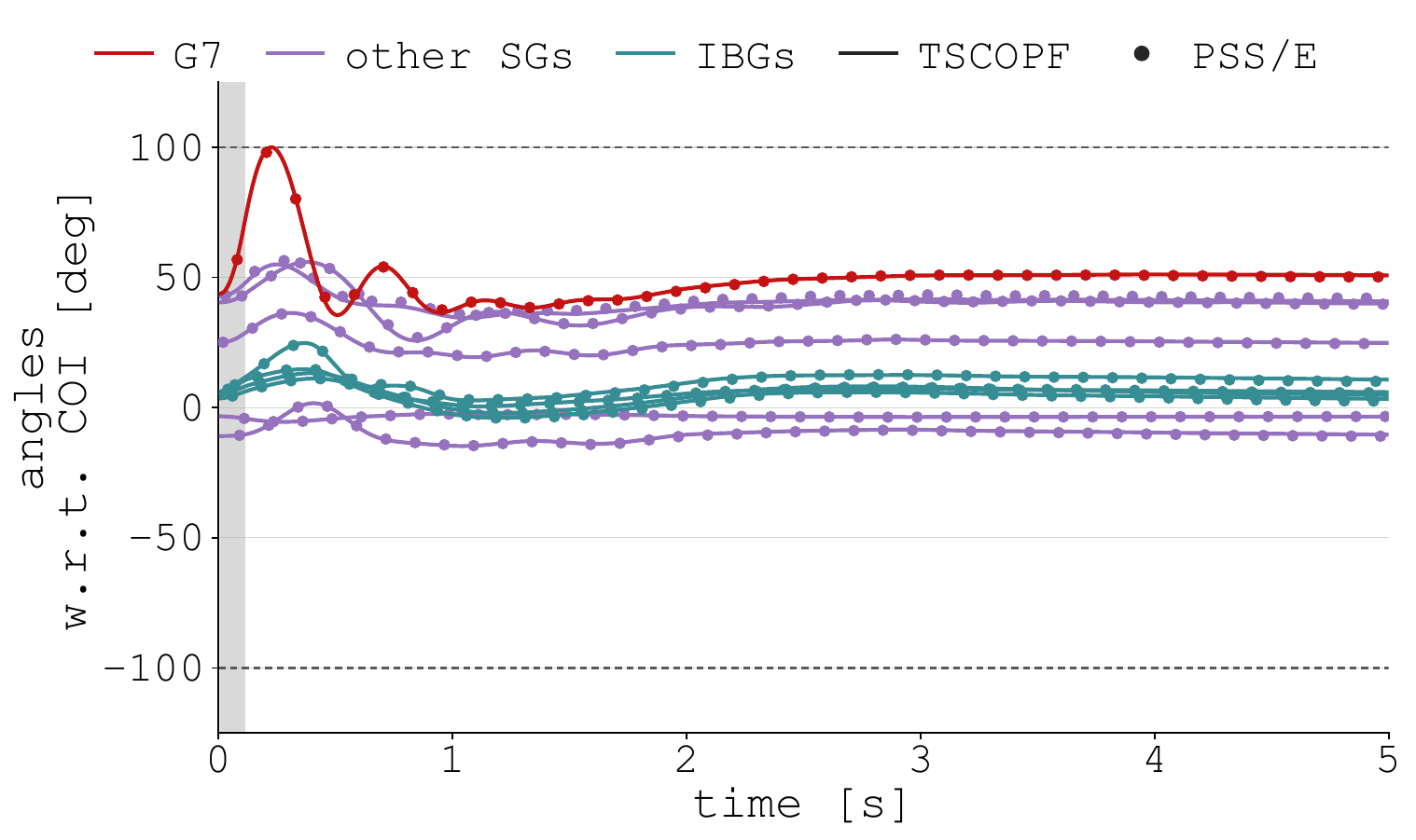}
    \caption{Angles relative to the \ac{COI} of the ten units obtained with \ac{TSCOPF} and with \texttt{PSS/E} for the short circuit at bus 23. The shaded area represents the short circuit before being cleared.}
    \label{fig:angles_39bus}
\end{figure}

Fig.~\ref{fig:angles_39bus} shows the angles of the ten units relative to the \ac{COI} obtained with \ac{TSCOPF} and compares them with those of \texttt{PSS/E}, a commercial time-domain simulation tool. In \texttt{PSS/E}, the \ac{TSCOPF} dispatch is set as the pre-fault operating point and the same contingency is simulated with the same dynamic data. The angles of all units closely match, during the short circuit (shaded area) and after its clearance. This agreement indicates that the discretized \acp{DAE} embedded in \ac{TSCOPF} reproduce the dynamics of a conventional time-domain simulation, for both \acp{SG} and \acp{IBG}. The figure also shows that a single stability constraint is active, the upper limit \eqref{eq:ineq_angle_coi_max} of G7 at one time step, whereas the four \acp{IBG} remain within $25^\circ$ of the \ac{COI}.

\begin{table}[!tb]
\captionsetup{justification=centering, textfont={sc,footnotesize}, labelfont=footnotesize, labelsep=newline}
\centering
\caption{Dispatch and prices of each generator in \ac{ACOPF} and in \ac{TSCOPF} for the 39-bus system. Generators are named as in Fig.~\ref{fig:39bus_system}.}
\label{tab:prices_39bus_gfm}
\footnotesize
\setlength{\tabcolsep}{2pt}
\begin{tabular}{llr>{\hspace{5pt}}rr>{\hspace{5pt}}rrrr}
\toprule
 & & & \multicolumn{2}{c}{\acs{ACOPF}} & \multicolumn{4}{c}{\acs{TSCOPF}} \\
\cmidrule(lr){4-5}\cmidrule(lr){6-9}
 & Unit & $\mathrm{c}_g$ & $P$ & $\lambda_k$ & $P$ & $\lambda_k$ & $\psi_{g(k)}$ & $\pi_{g(k)}$ \\
 & & {\scriptsize [\$/MWh]} & {\scriptsize [MW]} & {\scriptsize [\$/MWh]} & {\scriptsize [MW]} & {\scriptsize [\$/MWh]} & {\scriptsize [\$/MWh]} & {\scriptsize [\$/MWh]} \\
\midrule
\multirow{6}{*}{\acs{SG}} & G1 & 6 & 759.9 & 6 & 686.3 & 5.72 & $+$0.28 & 6 \\
 & G2 & 14 & 650 & 46 & 650 & 26.58 & $+$0.17 & 26.75 \\
 & G7 & 18 & 676.1 & 18 & 606.8 & 41.38 & $-$23.38 & 18 \\
 & G8 & 31 & 0 & 12.53 & 0 & 11.66 & $+$0.37 & 12.03 \\
 & G9 & 22 & 850 & 29.28 & 850 & 26.97 & $+$0.28 & 27.25 \\
 & G10 & 27 & 481.7 & 27 & 624.7 & 25.81 & $+$1.19 & 27 \\
\midrule
\multirow{4}{*}{\acs{IBG}} & G3 & 1 & 750 & 46.94 & 750 & 33.46 & $+$0.04 & 33.50 \\
 & G4 & 1 & 635.9 & 1 & 636.3 & 1.06 & $-$0.06 & 1 \\
 & G5 & 1 & 550 & 1 & 550 & 1.05 & $-$0.02 & 1.03 \\
 & G6 & 1 & 550 & 11.48 & 550 & 41.63 & $-$0.64 & 40.99 \\
\bottomrule
\end{tabular}
\end{table}

This case examines whether the properties observed in Section~\ref{subsec:9bus_system} hold in a network with losses, congestion and \acp{IBG}. Table~\ref{tab:prices_39bus_gfm} presents the dispatch and the prices for each generator. The \ac{TS} constraints increase the operating cost by $3.6\%$. The single active constraint, that of G7, yields a stability component for every asset of the system. For an \ac{IBG}, it follows from the initialization of its own dynamic model, as anticipated in Section~\ref{subsec:dynamic_model}, which shows that the derivation of Section~\ref{sec:economics_tscopf} does not depend on the generation technology.

The stability component separates G7 from the assets connected around it, which a branch limit cannot do. In \ac{ACOPF}, line 21--22 operates at its limit and lowers the nodal energy price of every asset behind it, including G6 and G7. In \ac{TSCOPF}, the output of G7 is restricted by its stability limit, the line is relieved, and the nodal energy components at both buses rise to $41.63$ and $41.38$~\$/MWh. The two units are nevertheless paid differently. G7 carries a stability component of $-23.38$~\$/MWh and is paid $\pi_{g(k)}=\mathrm{c}_g=18.00$~\$/MWh, whereas G6 carries only $-0.64$~\$/MWh and is paid $40.99$~\$/MWh. Paid $\lambda_k$ alone, G7 would receive $23.38$~\$/MWh above its marginal cost, although it is the unit that pushes the system toward the stability limit and whose output is reduced to maintain stability. The shadow price $\pi_{g(k)}$ removes this inconsistency, as discussed in Section~\ref{subsec:price_gen}, and charges the cost of securing stability to the asset that drives the active constraint.

The relation between $\lambda_k$ and the marginal cost established for the lossless network also holds here. G1 and G10 improve the stability margin and operate within their output limits, with $\lambda_k$ below $\mathrm{c}_g$ by $0.28$ and $1.19$~\$/MWh, whereas $\lambda_k$ exceeds $\mathrm{c}_g$ for G7. At the shadow prices, the three units recover their operating cost. On the demand side, the stability components do not exceed $1.02$~\$/MWh in magnitude and represent less than $0.1\%$ of the load payments, which confirms that the stability-related charges fall almost entirely on G7. 

Table~\ref{tab:settlement_39bus_gfm} compares the resulting settlement with that of \ac{ACOPF}. The residual decreases from $92.34$ to $84.13$~k\$/h. Relieving line 21--22 reduces the congestion collected in the residual by more than the stability components add to it. 

Overall, the results obtained in the 39-bus system confirm that the stability component is specific to each asset. They also show that the amount collected through the stability components adds to the residual that losses and branch congestion already produce through the nodal energy components. Moreover, the hybrid generation fleet, which includes \ac{IBG} models, shows that the stability components are obtained regardless of the dynamic model adopted for each generator, which indicates that they follow from the formulation itself.

\begin{table}[!tb]
\captionsetup{justification=centering, textfont={sc,footnotesize}, labelfont=footnotesize, labelsep=newline}
\centering
\caption{Operating cost and settlement at the shadow prices in \ac{ACOPF} and in \ac{TSCOPF} for the 39-bus system.}
\label{tab:settlement_39bus_gfm}
\footnotesize
\setlength{\tabcolsep}{4pt}
\begin{tabular}{lrrrr}
\toprule
 & Cost & Load payments & Generator revenues & $R$ \\
 & [k\$/h] & [k\$/h] & [k\$/h] & [k\$/h] \\
\midrule
\acs{ACOPF} & 60.02 & 219.57 & 127.23 & 92.34 \\
\acs{TSCOPF} & 62.19 & 205.46 & 121.33 & 84.13 \\
\bottomrule
\end{tabular}
\end{table}

\section{Conclusions}
\label{sec:conclusions}

The inclusion of transient stability constraints in the dispatch modifies the prices used to settle the market, an effect that had not been characterized for individual assets. To address this gap, shadow prices associated with generators and loads have been derived from a \ac{TSCOPF} in which the \acp{DAE} are embedded through direct discretization. Each shadow price has been shown to comprise a nodal energy component and a stability component, the latter given by the sensitivity of the active stability constraints to the power of the asset, weighted by their dual variables.

The case studies have demonstrated that the stability component is specific to each asset. Co-located generators, and co-located loads with different transient behaviors, are subject to different shadow prices, which a single \ac{LMP} cannot reflect. Even in a lossless and uncongested network, the nodal energy component has been shown to lie below the marginal cost of the generators that improve the stability margin and above that of the generators that reduce it. The proposed shadow prices let every generator operating within its output limits recover its operating cost, and allocate the cost of securing transient stability among the assets without a separate allocation rule, charging those that reduce the stability margin and remunerating those that improve it. In the 39-bus system, which includes losses, congestion and a hybrid fleet comprising \acp{SG} and \acp{IBG}, a single stability constraint is active, and the stability component of the generator that reaches it has been found to exceed that of any other asset by more than one order of magnitude.

Although the proposed shadow prices determine what each asset pays or receives, they leave a settlement residual that arises even in the absence of losses and congestion. As future work, the allocation of this residual should be addressed. Flexible demand is an extension of the formulation, in which a price-responsive load would react to its shadow price. Extending the analysis to commitment decisions is also relevant, as inertia is held fixed here and the binary variables involved prevent obtaining dual variables for pricing. Finally, the computational performance for sets of critical contingencies remains to be assessed in systems of realistic size.

\bibliographystyle{IEEEtran}
\bibliography{references}


\end{document}